\documentclass[final]{IEEEtran}

\usepackage[noadjust]{cite}

\usepackage{newtxtext,newtxmath}

\usepackage{graphicx}
\usepackage[switch]{lineno}
\usepackage[colorlinks=true, allcolors=blue]{hyperref}
\usepackage{upgreek}

\begin{document}
\title{A Multi-Function Radiation-Hardened HV and LV Linear Regulator for SiPM-based HEP Detectors}
\author{Paolo Carniti, Claudio Gotti, and Gianluigi Pessina
\thanks{The authors are with INFN Milano-Bicocca and University of Milano-
Bicocca, Piazza della Scienza 3, Milano, 20126 Italy.\\
e-mail: paolo.carniti@mib.infn.it}}

\maketitle


\begin{abstract}
The use of silicon photomultipliers (SiPMs) to detect light signals in highly radioactive environments presents several challenges, particularly due to their sensitivity on radiation, temperature, and overvoltage, requiring a proper management of their bias supply.
This article presents the design and performance of ALDO2, an application-specific integrated circuit and power management solution tailored for SiPM-based high-energy physics detectors.
The chip's functions include adjustable and point-of-load linear regulation of the SiPM bias voltage (10-70~V, 50~mA), monitoring of SiPM current, shutdown, over-current and over-temperature protection.
The same functions are also available for the low-voltage regulator (1.6-3.3~V, 800~mA), used to generate the power supply of SiPM readout chips that often demand stable and well-filtered input voltages while consuming currents of up to several hundred milliamperes.
The chip is intended to operate in radioactive environments typical of particle physics experiments, where it must withstand significant levels of radiation (total ionizing dose and 1-MeV-equivalent neutron fluence in the range of Mrad and $\mathbf{10^{14}}\ \mathrm{\mathbf{n_{eq}/cm^2}}$, respectively).
The article provides a comprehensive description of the chip design, as well as experimental measurements, offering insights into the chip's performance under various conditions.
Finally, radiation hardening, radiation qualification and reliability are discussed.
	
\end{abstract}


\section{Introduction}
\label{sec:introduction}

\IEEEPARstart{S}{ilicon} photomultipliers (SiPMs, also known as multi-pixel photon counters or MPPCs)~\cite{Gundacker_2020} have gained wide-spread use in various high-energy physics (HEP) experiments, primarily due to their excellent photon counting performance, compactness, and immunity to magnetic fields.
Despite being more practical and durable than vacuum-based photosensors, integrating SiPMs into experiments still presents several challenges, particularly concerning their power supply.
Radiation damage, in fact, leads to a significant increase in the dark current, reaching up to a few milliamperes per device, and a shift in the breakdown voltage, which can deviate by several hundred millivolts from the typical value of a few tens of V.
To mitigate the effects of radiation, SiPMs are typically cooled below $0\,^\circ$C~\cite{GARUTTI201969, CALVI2019243}, adding further drifts due to temperature variations.
All these shifts in the breakdown voltage, in addition to the intrinsic production spread, must be compensated by the bias supply generator throughout the experiment lifetime, since SiPM performance depends strongly on the overvoltage (the excess bias over the breakdown voltage).

The ALDO2 is an application specific integrated circuit (ASIC) specifically designed to handle the power management of detectors using SiPMs (usually grouped in arrays of 8 to 16 devices).
Its primary functions are summarized in the following:
\begin{itemize}
	\item Adjustable regulation of the SiPM bias voltage
	\item Point-of-load (PoL) regulation of SiPM bias voltage to enhance stability and noise filtering
	\item Monitoring of SiPM current for on-detector I-V curve characterization
	\item Shutdown, over-current and over-temperature protection
	\item PoL regulation and filtering for the low voltage supply of front-end ASICs
\end{itemize}

The chip has to operate in highly radioactive environments, typical of HEP experiments.
Given the chip proximity to the SiPMs, the maximum radiation levels are constrained by the capabilities of currently available SiPM models, specifically a 1-MeV-equivalent neutron fluence of a few $10^{14}\ \mathrm{n_{eq}/cm^2}$, and a total ionizing dose (TID) of a few Mrad.

\section{Chip overview}
\label{sec:overview}

ALDO2 is fabricated using onsemi I3T80 0.35~$\upmu$m CMOS technology. 
I3T80 was specifically selected since it provides both low-voltage (LV) and high-voltage (HV) MOSFETs, as well as adequate radiation tolerance~\cite{FaccioHVradhard}.
Only devices with thin gate oxide and maximum $\mathrm{V_{gs}}$ of 3.3~V are used in the ASIC as they are the only ones suitable for the target radiation levels.
I3T80 includes lateral and vertical double-diffused MOSFETs (DMOSFET) that can operate with $\mathrm{V_{ds}}$ up to 70~V, as well as floating wells, diodes, matched resistor arrays and bipolars.
Both LV and HV n-type devices require specific custom layout modifications to attain the target radiation tolerance, as detailed in Sec. \ref{sec:layout}.
It is also known that DMOSFETs experience an increase in on-resistance as a result of displacement damage in the drift regions, although this effect is expected to become significant for I3T80 only beyond the specified radiation levels~\cite{FaccioHVradhard}.


\begin{figure}[b]
	\centering
	\includegraphics[height=3.15cm]{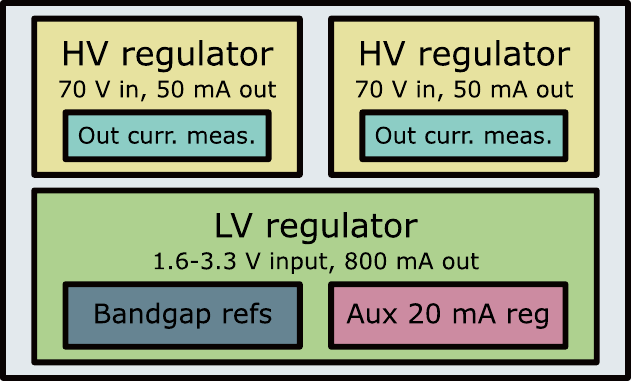}
	\hfill 
	\includegraphics[height=3.15cm]{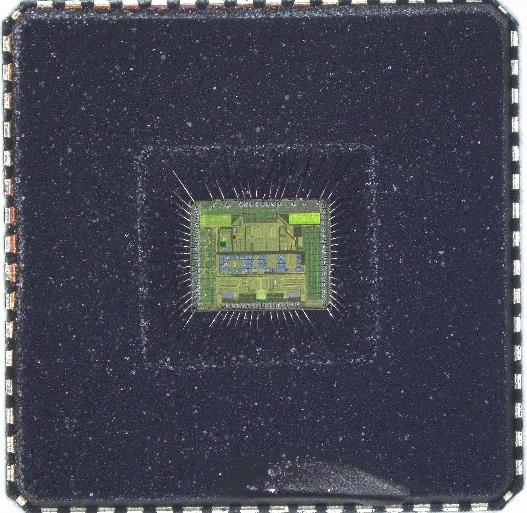}
	\caption{General block schematic of ALDO2 (left). Photograph of a decapsulated ALDO2 in QFN64 package (right). Die dimensions are $\mathrm{2.5\times2\,mm^2}$.}
	\label{fig:block_sch}
\end{figure}

The chip consists of two completely independent sections, HV and LV, as illustrated in the general block schematic in Fig.~\ref{fig:block_sch}.
A photograph of the chip is also shown.

\begin{figure}
	\centering
	\includegraphics[width=.93\columnwidth]{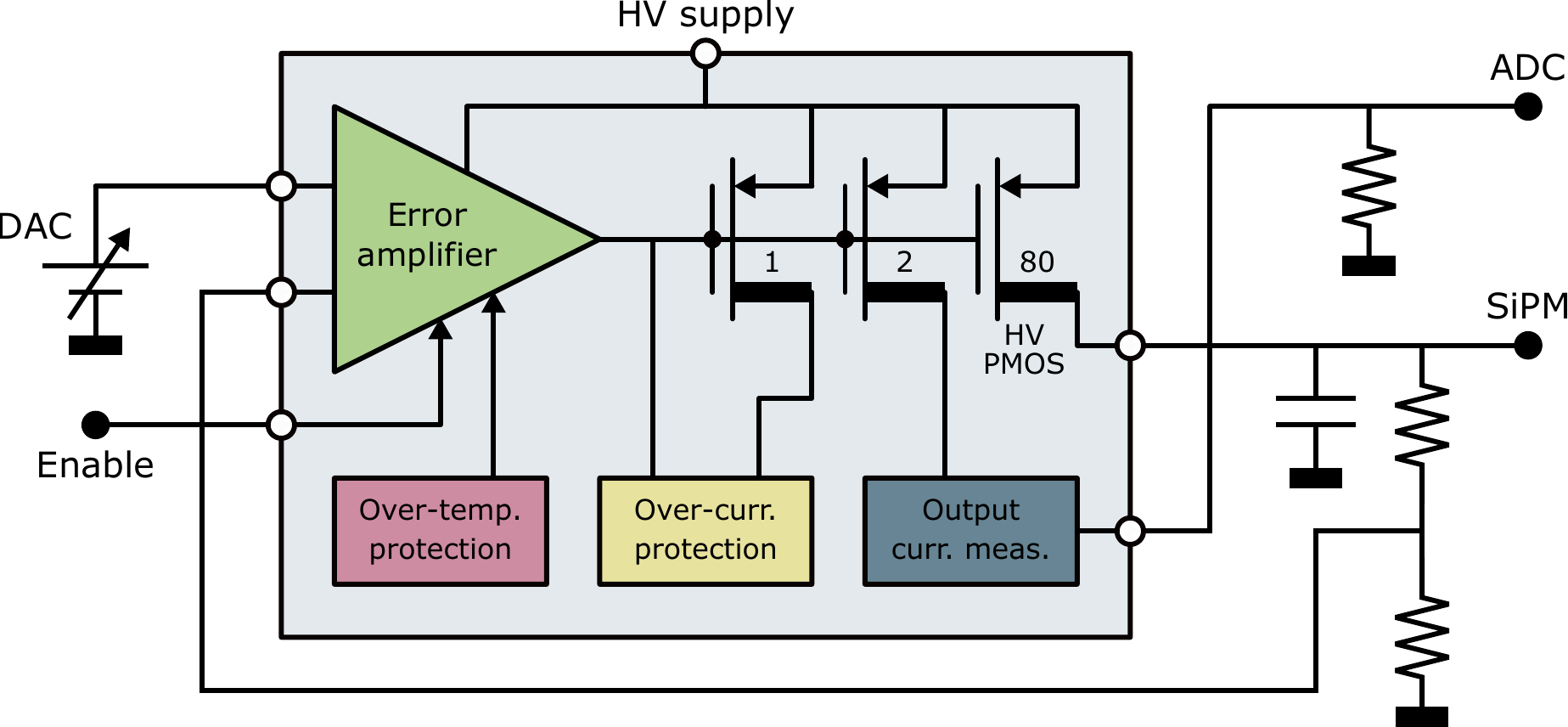}
	\caption{Simplified block schematic of the HV regulator (two identical ones are included in each ALDO2). Devices with thick drain are HV DMOSFETs.}
	\label{fig:block_sch_hv}
\end{figure}

The HV section comprises two identical low drop-out (LDO) linear regulators.
Each one can regulate an input voltage of maximum 70~V down to an adjustable value, and deliver up to 50~mA to the load (an array of SiPMs).
The block schematic for this section is presented in Fig.~\ref{fig:block_sch_hv}.
During the preliminary design phase, various topologies were considered. Ultimately, high-side active linear regulation was deemed the most suitable, as it offers superior stability and noise characteristics compared to passive trimming, while maintaining $>$90$\,\%$ efficiency (provided that dropout is in the $\sim$V range).

To adapt the SiPM bias voltage during detector operation, the output voltage of the regulator is adjusted by applying a programmable voltage to the reference input of the regulator using an external DAC.
The DAC is integrated in the SiPM front-end ASIC or in other slow-control ASICs present in the system.
Typically, the DAC adjustment range is from 0.75 V to 1 V, although values up to 3.3 V can be used.
The feedback resistors are external, and the gain is usually between 30 to 50 V/V, depending on SiPM specifications.
The regulator can be disabled with a digital signal, useful in case of a malfunctioning SiPM.
It is also protected by an output current limiting circuit (without foldback), which operates alongside an over-temperature block that disables the error amplifier when the die temperature exceeds a fixed threshold.
The output current is measured by copying it through current mirrors, then converted to a voltage using an external resistor, and finally fed to an external ADC.

\begin{figure}
	\centering
	\includegraphics[width=.93\columnwidth]{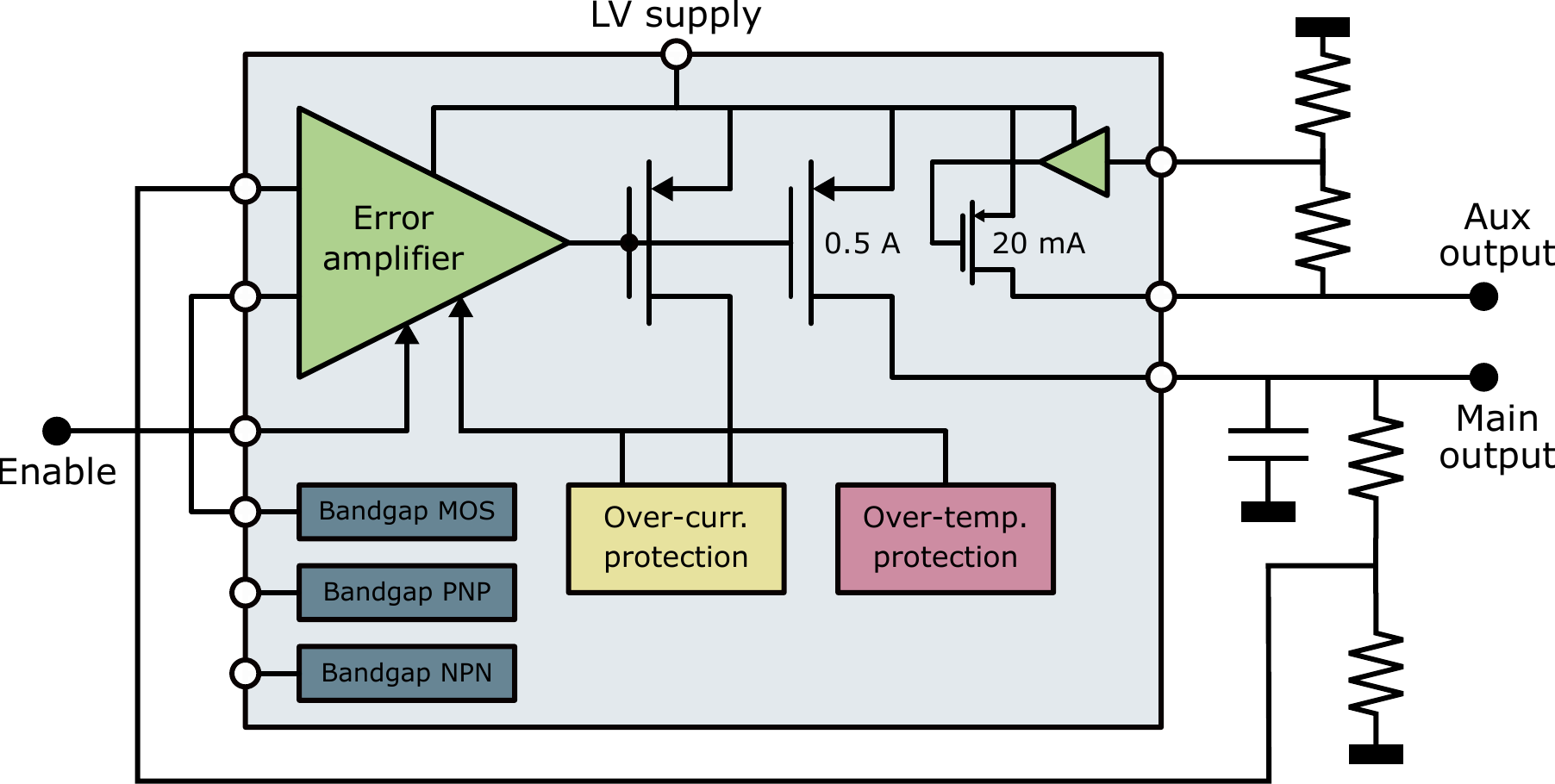}
	\caption{Simplified block schematic of the LV regulator.}
	\label{fig:block_sch_lv}
\end{figure}

The LV section incorporates a high-current (maximum 0.8~A) primary LDO regulator, a low-current (20~mA) auxiliary LDO regulator, and integrated bandgap voltage references, as illustrated in the block schematic in Fig.~\ref{fig:block_sch_lv}.
This section's design is similar to ALDO2 predecessor, ALDO1~\cite{aldo1}.
Both the main and auxiliary LDOs are adjustable using external feedback resistors.
The LV section adopts only LV devices, thereby limiting the maximum input supply to 3.3~V.
A linear topology is preferred to achieve the best possible filtering of the input supply, which is typically generated by DC-DC regulators.
Similarly to the HV section, the regulators are protected with over-current and over-temperature circuits.
Three bandgap voltage reference circuits are included, each based on the same topology but with different devices to generate the PTAT/CTAT voltages (PNP bipolars, NPN bipolars, and dynamic-threshold MOS transistors or DTMOST~\cite{assaderaghi1994dtmos}).
The bandgap selection depends on the final application and is made externally by connecting the bandgap output to the voltage reference input of the error amplifier.
Bipolar-based bandgaps are expected to have better thermal stability and uniformity. DTMOST-based one has higher radiation tolerance and lower reference voltage (0.63~V).

The chip is packaged in QFN64 ($\mathrm{9\times9\,mm^2}$) and has been produced in large quantities (45k chips from 26 wafers) for the use in two detectors of the CMS experiment at CERN, the Barrel Timing Layer~\cite{CERN-LHCC-2019-003, Abbott_2021} and the High-Granularity Calorimeter~\cite{CERN-LHCC-2017-023}.

\section{High voltage section}
\label{sec:hv}

\subsection{Circuit description}
\label{sec:hv_circuit}

\begin{figure*}
	\centering
	\includegraphics[width=.93\textwidth]{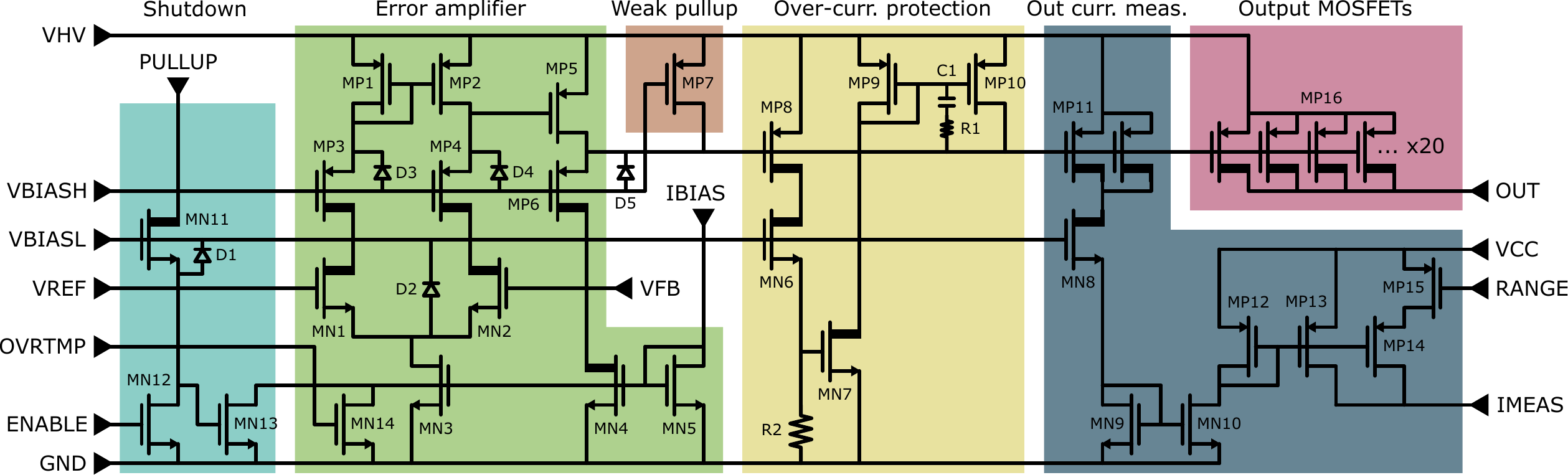}
	\caption{Full schematic of the HV regulator. All MOSFETs have the substrate connected to their source. Devices with thick drain are HV DMOSFETs. Diodes are made with diode-connected MOSFETs.}
	\label{fig:sch_hv}
\end{figure*}

The core parts of ALDO2 are the two HV LDO regulators. 
Fig.~\ref{fig:sch_hv} shows their complete schematic.
The regulator is required to be powered between the HV rail and ground, while using only thin-oxide transistors with a maximum gate voltage of 3.3~V.
The p-type pass element and its driving circuitry operate close to the positive rail to minimize the dropout, while the input pair and other circuits operate close to the ground rail since they must interface with external LV components (SiPM readout ASIC, DAC, ADC).
More details are discussed in the following.

The error amplifier (green block) is a two-stage n-type-input opamp. 
The circuit employs a voltage limiting solution using cascode p-type DMOSFETs (MP3, MP4, and MP6) with gates biased by the external voltage VBIASH (2-3~V below the HV rail) to protect regular LV p-type MOSFETs (MP1, MP2 and MP5).
The role of these DMOSFETs is twofold:
they can withstand the full HV drop between source and drain in case of significant input pair imbalance;
they limit the drop across LV p-type MOSFETs in various conditions where the drain of the LV PMOS would be pulled below  $\mathrm{V_{BIASH}+V_{th,pDMOS}}$ (such as saturation or off-state).
A similar approach is commonly used in level shifters for digital circuits and DCDCs~\cite{le1998high}.

The concern of exceeding the maximum rating also extends to the input pair (MN1-2), whose drain voltage can range up to the positive rail in normal use (weak or moderate saturation).
For this reason, MN1-2 are implemented with HV n-type vertical DMOSFETs.
The main drawback of this choice is that some electrical parameters (e.g., transconductance, matching, etc.) are worse than standard LV n-type devices.
Nevertheless, the random offset (evaluated with Monte Carlo simulation over process and mismatch) is below 0.5~mV RMS.
An alternative solution involving LV input MOSFETs, protected using a similar concept as the one used for the active load, was considered but ultimately rejected due to area occupation.


MN1-2 have a width of 500~$\upmu$m, while their length is fixed by technology.
The protection p-type DMOSFETs also have fixed channel length, and a width of 40~$\upmu$m.
The active load has dimensions (in $\upmu$m) of 4800:1, the second gain stage (MP5) 800:0.35.
The input pair is biased with 370~$\upmu$A, the second stage with  160~$\upmu$A, with a total current consumption of 730~$\upmu$A (35~mW at 48~V supply).

The regulator can be disabled by the shutdown block (light blue), which sinks the bias current of the error amplifier if the ENABLE signal is above the MN12 threshold.
Transistors MN11-13 implement the default-off condition when the ENABLE is pulled to ground, without requiring an additional power rail apart from the HV one.
MN11 is used as a voltage limiting element, similarly to MP3-4 of the error amplifier, with VBIASL set to 2-3~V above ground.
An off-chip 10~M$\Omega$ resistor is connected to PULLUP.
The OVRTMP signal from the over-temperature protection circuit functions in a similar manner to the shutdown circuit.
The default state of the regulator (off) is also granted by MP7, which pulls the output of the error amplifier to the HV rail when the bias current is interrupted by the shutdown block.

Diodes D1-5 ensure that no sensitive nodes are driven below (above) VBIASH (VBIASL) by more than a diode drop due to leakage of off-state voltage-limiting DMOSFETs.

The output pass element of the LDO regulator is a p-type HV DMOSFET (MP16, magenta block) made up of 80 fingers, with a total width of 2000~$\upmu$m.

The over-current protection circuit (yellow block) operates by mirroring the output current with a 1:80 ratio (MP8) and then comparing the voltage drop of the mirrored current (developed over the internal resistor R2, 700~$\Omega$) with the threshold voltage of transistor MN7.
If the output current exceeds approximately 55~mA, MP9 and MP10 pull up the output of the error amplifier, causing the regulator to enter a regime of constant output current.

The output current measurement circuit, or OCM (dark blue block), follows a similar concept by mirroring (i.e., attenuating) the output current with a 1:40 ratio (MP11), then a 1:1 ratio (MN9-10), and finally a programmable (with RANGE pin) 1:1 or 1:20 ratio (MP12-13/14), achieving a total attenuation ratio of 1:40 or 1:800.
The mirrored output current can be converted to a voltage with an external resistor of the desired value and read with an external ADC.
The final p-type mirroring stage is powered by a LV power rail (VCC), the same as the ADC.

Voltages VBIASH and VBIASL can either be internally generated using a series of diode-connected MOSFETs, or applied externally by using a voltage divider between the HV rails.
The former method is preferred as it ensures more stable voltages when changing the input HV supply, thereby providing a wider input supply range.

The ALDO2 employs the standard technique for LDO stabilization with an external low-ESR tantalum capacitor connected to the output node~\cite{rinconmora1998ldostab}.
Polymer tantalum capacitors do have the required radiation hardness~\cite{martin2016radiation}, but are limited in terms of voltage rating (50~V or 63~V), capacitance (below 10~$\upmu$F) and ESR (about 100~m$\Omega$).
The design had to cope with these constraints.

The dominant pole of the LDO is the one at the output node, determined by the output capacitance (few $\upmu$F) and the output resistance, which consists of the parallel combination of the output PMOS $\mathrm{R_{DS}}$ (9.7~k$\Omega$), the feedback resistors (about 200~k$\Omega$), and the load resistance (1~k$\Omega$ in the worst case).
Its frequency thus ranges between 10 and 100~Hz.
With a loop gain of 83~dB, the unity-gain bandwidth (UGB) is about 500~kHz.

The second pole in the feedback loop is at the output node of the error amplifier (gate of MP16).
The capacitance at that node is 5.5~pF, the output resistance is 100~k$\Omega$, and the pole frequency is 290~kHz (smaller than the UGB).
However, the low-ESR capacitor stabilizes the loop by introducing a zero at a frequency of $\mathrm{1/\left(2\pi C_O R_{ESR}\right)}$.
The selected capacitor (Kemet T521, rated 50~V, 1.8~mm thick) has $\mathrm{C_O}$ of 10~$\upmu$F and $\mathrm{R_{ESR}}$ of 90~m$\Omega$, resulting in a zero at 180~kHz, which cancels the second pole in any operating corner.
Other poles are well above the UGB.

When the over-current protection is engaged, a local loop from MP8 to MP10 is also active.
C1 and R1 (3~pF and 3~k$\Omega$) compensate the regulator in this condition.

The RMS output noise (1~Hz to 10~MHz) is 370~$\upmu$V, with the dominant contribution at the typical maximum load (20~mA) originating from the input pair.
It is important to note that due to the relatively high gain of the regulator (30-50 V/V), the contribution of the DAC noise can be even more significant.
Therefore, ensuring proper design of the DAC, along with adequate RC filtering, is essential for optimal performance.

The power supply rejection (PSR) is better than -70~dB from DC to 1 kHz, thanks to the high loop gain of the regulator.
It reaches a minimum of -28~dB at 200~kHz, where the path through the gate capacitors of the PMOS pass transistor becomes most relevant.



%
%

\subsection{Experimental measurements}
\label{sec:hv_meas}

The performance of the ALDO2 ASIC has undergone comprehensive laboratory qualification across various operating conditions (input and output voltages, load, temperature, etc.).
Preliminary measurements have been previously presented~\cite{aldo2022}.

\begin{figure}
	\centering
	\includegraphics[width=.48\columnwidth]{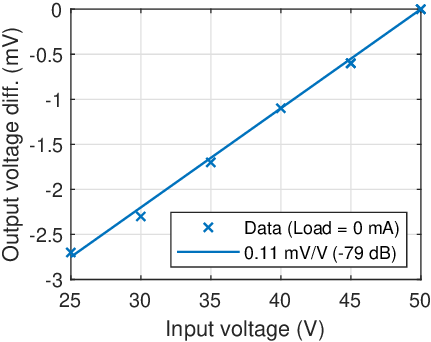}
	\hfill 
	\includegraphics[width=.48\columnwidth]{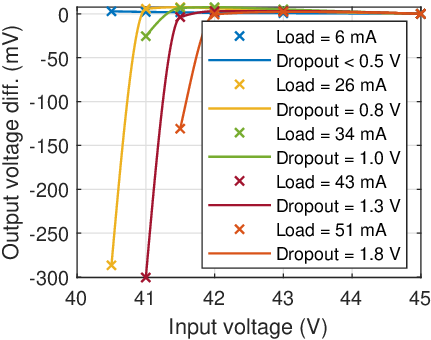}
	\caption{Left: Line regulation (DC PSR) with no output load and output voltage set at 20~V. Right: Measurement of the minimum dropout at different loads with the output voltage set at 40~V.}
	\label{fig:hv_linereg_dropout_noload}
\end{figure}

In Fig.~\ref{fig:hv_linereg_dropout_noload}, the left plot depicts the line regulation (DC PSR) measurement with no load current aside from that through the feedback network (100~$\upmu$A) and the output voltage set at 20~V.
The measured value of -79~dB matches the simulation.
The plot on the right shows the measurement of the minimum dropout, ranging from below 0.5~V at 6~mA load to 1.8 V at 51~mA.
The ALDO2 achieves a load regulation of 0.015$\,\%$ (6~mV change at 40~V output) with a load current step of 20~mA, or 7.5~ppm/A.

\begin{figure}[b]
	\centering
	\includegraphics[width=.49\columnwidth]{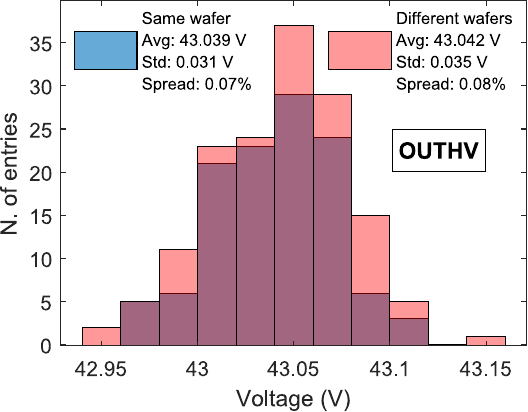}
	\hfill 
	\includegraphics[width=.49\columnwidth]{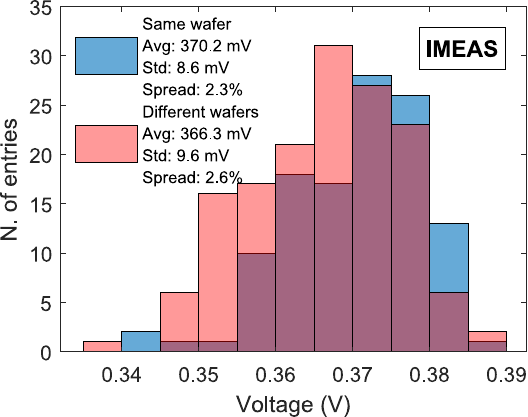}
	\caption{Distributions of the HV regulator output voltage (OUTHV, left) and output current measurement (IMEAS, right) for 117 chips on the same wafers (blue) and 152 chips randomly picked from 8 different wafers (red).}
	\label{fig:hv_spread}
\end{figure}

The left plot of Fig.~\ref{fig:hv_spread} shows that the RMS spread of the output voltage is 0.07$\,\%$ (0.7 mV input-referred), highlighting that the contribution of the error amplifier (due to offset and finite gain) is negligible compared to the spread of the feedback resistors and DAC, which are typically 1$\,\%$.
The two distributions represent samples of chips from the same wafer (blue) or randomly picked from different wafers (blue).
No appreciable difference was measured between the two.
The histograms on the right plot show the spread of the OCM output (with 20~mA load), amounting to 2.6$\,\%$ in samples from different wafers.

\begin{figure}
	\centering
	\includegraphics[width=.49\columnwidth]{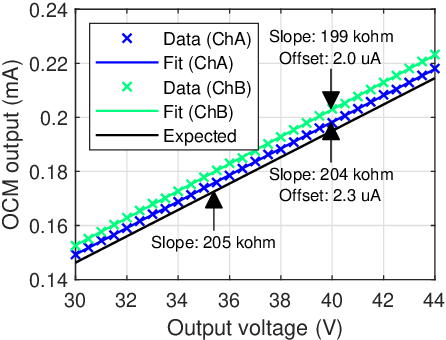}
	\hfill
	\includegraphics[width=.49\columnwidth]{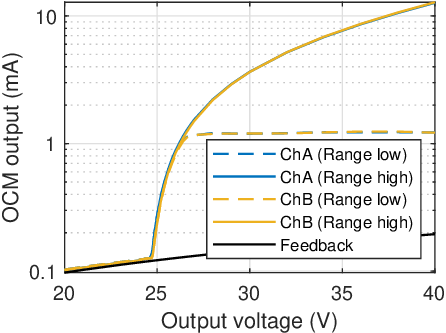}
	\caption{OCM characterization. Left: Only feedback resistors, low range. Right: With irradiated SiPM (SensL MicroFC), low range (solid line) and high range (dashed line). The low range saturates just above 1~mA.}
	\label{fig:hv_curr_meas}
\end{figure}

Fig.~\ref{fig:hv_curr_meas} presents the characterization of the OCM circuit.
The left plot shows the output current as a function of the output voltage with no load apart from the feedback resistors, exhibiting excellent linearity, as expected for a resistive load, and matching the ideal curve (black).
The range of the circuit is set to low (1:40 ratio).
The right plot demonstrates a practical application of the OCM circuit, measuring the I-V curve of an irradiated SiPM with an ADC.
The measured value is the sum of the feedback current and the dark current of the SiPM.
This property can be used to further calibrate the OCM circuit when radiation damage accumulates and possible drifts occurs.
By setting an output voltage lower than the breakdown of the SiPM, in fact, the only output current is due to the feedback, which is fixed and known from the value of the feedback resistors and of the output voltage (both not changing with radiation damage).

Noise measurements have confirmed the simulation results, indicating 350~$\upmu$V RMS noise at the typical maximum load.

\section{Low voltage section}
\label{sec:lv}

The low voltage section of ALDO2 comprises the two LDOs (primary and aux), the protection circuits, and the bandgap voltage references.

\subsection{Low voltage LDOs}
\label{sec:lv_ldo}

The two LV regulators (0.8~A and 20~mA ones) share identical designs, differing only in the width of output p-type MOSFET (the length is 0.35~$\upmu$m). The high-current LDO features a 32~mm wide one, while the low-current LDO has a 1~mm wide one.

\begin{figure}[b]
	\centering
	\includegraphics[width=\columnwidth]{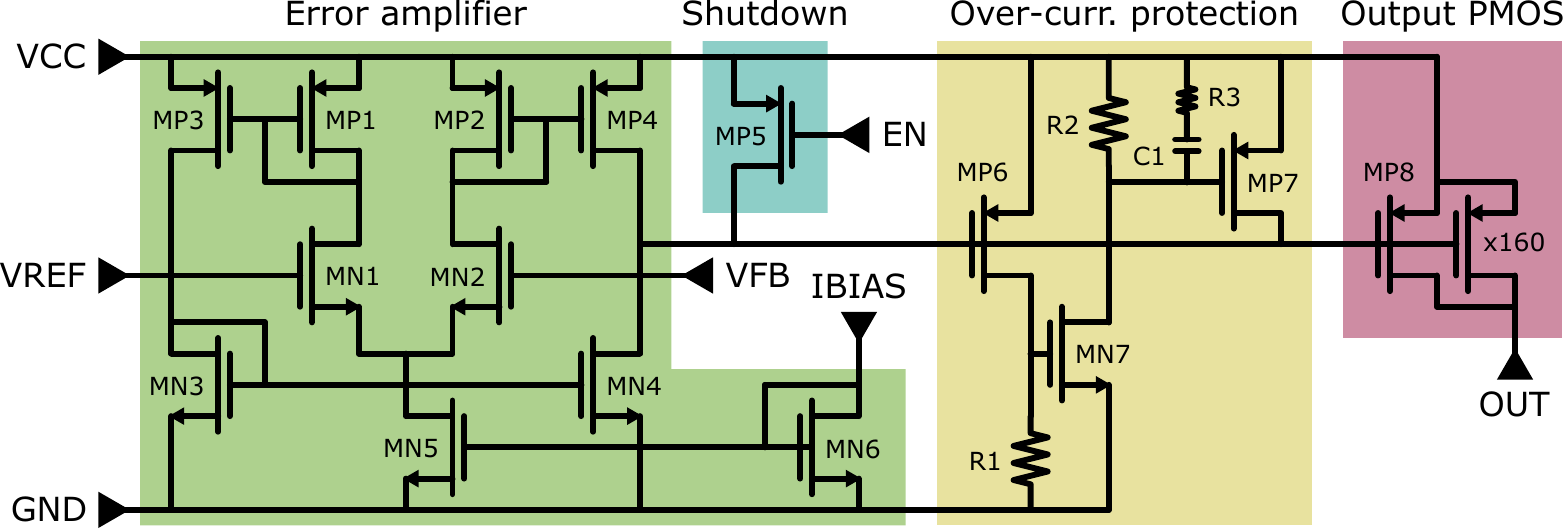}
	\caption{Full schematic of the LV regulator. All MOSFETs have the substrate connected to their source.}
	\label{fig:sch_lv}
\end{figure}

The complete schematic of the high-current regulator is presented in Fig.~\ref{fig:sch_lv}.
The error amplifier (green block) is a single-stage NMOS-input opamp with a mirrored topology and a gain of 44~dB.
The use of an n-type input pair is essential as the error amplifier needs to accommodate all the three types of bandgaps, each with different output voltages (1.2~V for the BJT-based ones and 0.63~V for the MOS-based one), while minimizing the input voltage (1.5~V at low loads).
A p-type input pair would not be suitable with BJT-based bandgaps, as the $V_{gs}$ of the input pair would be below 300~mV at 1.5~V supply.
This compromise, although mandatory, results in increased overall noise (due to the higher flicker noise of n-type devices) and worse matching when the regulator is used with the DTMOST-based bandgap (due to the operation of the input pair in moderate inversion).
The input pair (MN1-2) has a size of 1000:0.7, MP1-2 of 2000:5, MP3-4 of 1000:5, and MN3-4 of 1000:2 (sizes in $\upmu$m).
The mirrored topology provides a larger output swing of the error amplifier towards the ground rail, and thus larger overdrive capability for the output PMOS transistor.

The regulator can be externally disabled (light blue block) by pulling down the EN input.
The over-temperature protection (not shown) also operates in the same way .
The over-current protection circuit (yellow block) features a similar architecture as the HV one, but with the R2 resistor (40~k$\Omega$) instead of a p-type mirror and adopting a different compensation scheme (C1 2.4~pF, R3 10~k$\Omega$).
The sense resistor R1 is 200~$\Omega$.

Both LV regulators are stabilized with external tantalum capacitors.
In the case of the high-current regulator, the gain of the error amplifier is lower (44~dB) than the HV counterpart, and the gate capacitance of the 32-mm-wide pass transistor is significantly higher (45.4~pF).
At the maximum output current, the output impedance of the amplifier is 220~k$\Omega$, placing the second pole at 16~kHz.
The combination of $\mathrm{C_O}$ and $\mathrm{R_{ESR}}$ must introduce a zero approximately in the same range (e.g., 10~$\upmu$F and 250 m$\Omega$ or 100~$\upmu$F and 25 m$\Omega$).
The availability of LV tantalum capacitors is much larger than HV ones, so the specific model selection is not critical in this case.

\subsection{Bandgap voltage reference}
\label{sec:lv_bandgap}

\begin{figure}
	\centering
	\includegraphics[width=0.98\columnwidth]{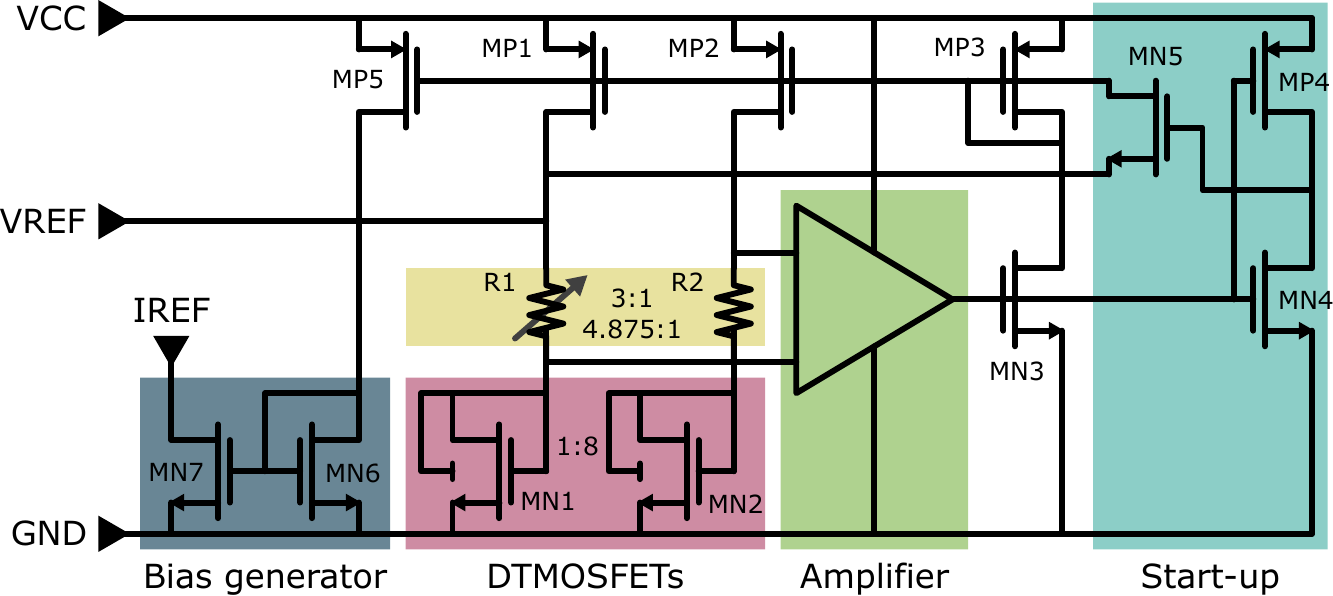}
	\caption{Schematic of the DTMOST-based bandgap. The BJT-based ones are idential, except for the resistor array, which is not trimmable. All MOSFETs have the substrate connected to their source (except MN1-2).}
	\label{fig:sch_bandgap}
\end{figure}

Fig.~\ref{fig:sch_bandgap} shows the schematic of the DTMOST-based bandgap, which adopts the modified Kuijk's topology presented in~\cite{annema1999bandgap}.
MN1-2 (red block) are the DTMOSTs used for generating the PTAT/CTAT voltages.
MN1 is sized 570:0.7 (in $\upmu$m), while MN2 is 8 times larger.
Unlike most CMOS technologies, I3T80 permits the implementation of DTMOSTs using n-type devices within isolated P-wells.
This feature is advantageous for radiation hardness due to the utilization of hardened layout for n-type devices, as discussed in Sec.~\ref{sec:layout}.

The resistor pair (yellow block) uses an array of matched polisilicon resistors.
The ratio of R1/R2 is externally trimmable to compensate for wafer or device spread, ranging between 4.875:1 and 3:1 with 4 bits (0.125 LSB).
The amplifier (green block) is a dual-stage compensated PMOS-input opamp.
The current mirrors that bias the DTMOSTs also generate the current reference (dark blue block) used in other parts of the LV section.

The bipolar-based bandgaps, either NPN and PNP, share most of the design, but the resistor pairs are not trimmable.


\subsection{Experimental measurements}
\label{sec:lv_meas}

\begin{figure}
	\centering
	\includegraphics[width=.48\columnwidth]{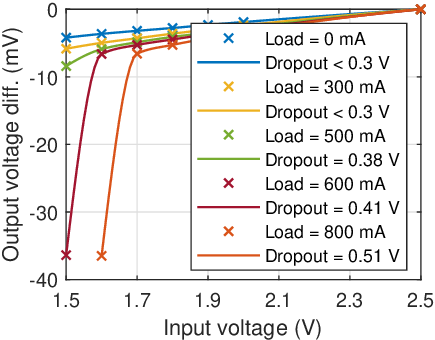}
	\hfill 
	\includegraphics[width=.48\columnwidth]{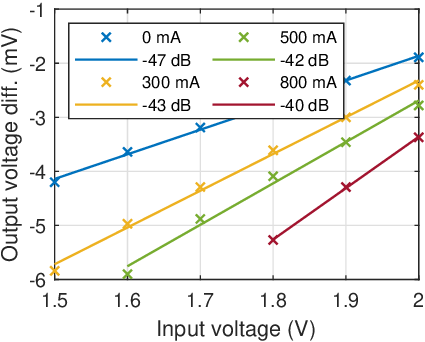}
	\caption{Left: Measurement of the minimum dropout at different loads and output voltage set at 1.2~V. Right: Zoom of the previous plot above the minimum dropout and measurement of the line regulation (DC PSR).}
	\label{fig:lv_linereg_dropout}
\end{figure}

Fig.~\ref{fig:lv_linereg_dropout} presents the measurements of the minimum dropout (left plot) and line regulation (right plot) at different loads.
The minimum dropout is defined as the dropout at which line regulation gets worse than -34~dB.
At the typical load of 500~mA, the minimum dropout is 0.4 V and the line regulation above that is -42~dB (8~mV/V).
The load regulation is 5 mV/A (0.42$\,\%$/A with 1.2 V output), evaluated using a current step of 0.5~A.

\begin{figure}[b]
	\centering
	\includegraphics[height=3.2cm]{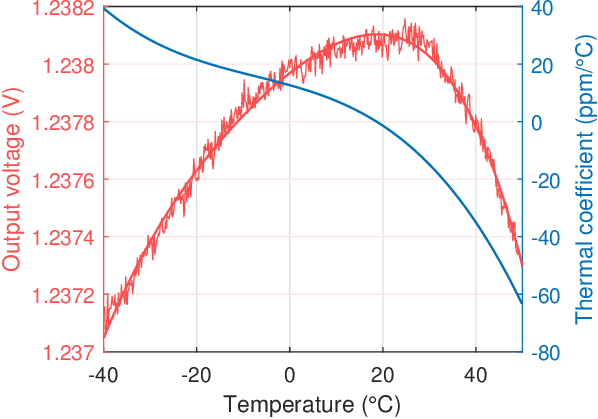}
	\hfill 
	\includegraphics[height=3.4cm]{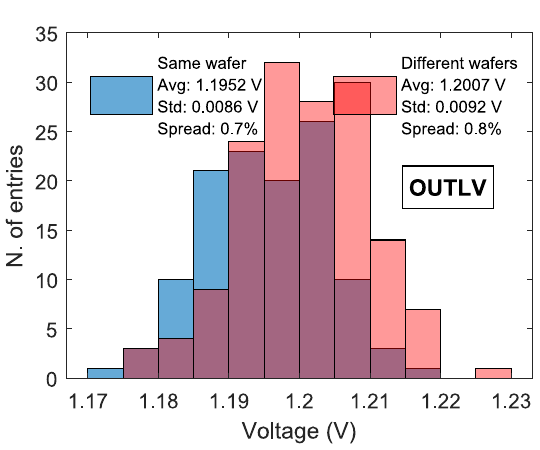}
	\caption{Left: Temperature drift of the LV output voltage while using DTMOST-based bandgap. Right: Distributions of the LV output voltage for 117 chips on the same wafers (blue) and 152 chips from 8 different wafers (red). All devices have the same bandgap trimming.}
	\label{fig:lv_drift_spread}
\end{figure}

Measurements of the bandgaps' thermal stability indicate that PNP-based one performs better than the others, exhibiting a maximum drift of 20~ppm/$^\circ$C in the temperature range from -40$\,^\circ$C to 80$\,^\circ$C.
The DTMOS-based bandgap displays a drift of 50~ppm/$^\circ$C and a higher quadratic slope.
The NPN-based one performs worse than expected, with a drift of about 120~ppm/$^\circ$C.
The thermal drift at the output of the regulator is dominated by the contribution of the bandgap and no other effects are introduced by the error amplifier.
The left plot of Fig.~\ref{fig:lv_drift_spread} shows the drift and thermal coefficient of the ALDO2 LV output voltage when the reference is provided by the DTMOS-based bandgap.

The right plot of Fig.~\ref{fig:lv_drift_spread} presents the distribution of the output voltage within the same wafer (blue data) or between different wafers (red data), highlighting an RMS spread of 0.7$\,\%$ and 0.8$\,\%$, respectively.
All these samples adopted the same trimming configuration of the bandgap.

\begin{figure}
	\centering
	\includegraphics[width=.8\columnwidth]{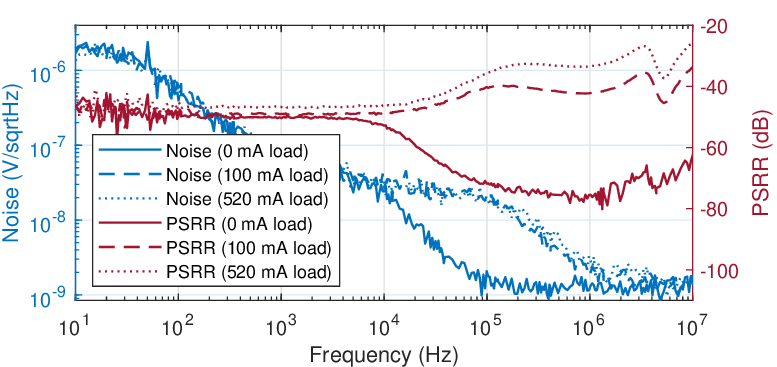}
	\caption{Noise (left axis) and PSRR (right axis) of the LV regulator at different output loads. Input and output voltages are 1.8~V and 1.2~V, respectively.}
	\label{fig:lv_noise_psrr}
\end{figure}

Fig.~\ref{fig:lv_noise_psrr} shows the measurement of the noise at the output of the regulator (left axis) at different output loads.
The RMS noise at 500~mA is 27~\textmu V (integrated from 10~Hz to 100~MHz).
The power supply rejection measurement (right axis) shows that ALDO2 is capable of rejecting input supply noise better than -40 dB up to 50~kHz and better than -26 dB up to 10~MHz.

\section{Over-temperature protection circuit}
\label{sec:overtemp}

\begin{figure}[b]
	\centering
	\includegraphics[width=0.84\columnwidth]{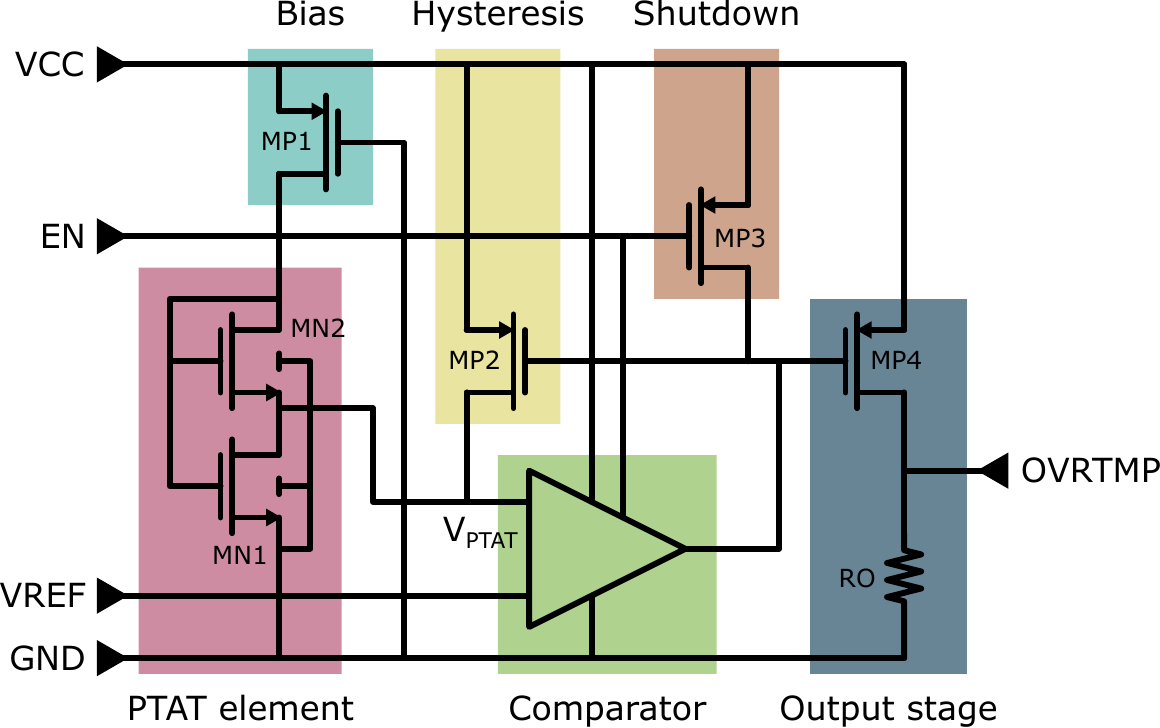}
	\caption{Schematic of the over-temperature protection circuit.}
	\label{fig:sch_overtemp}
\end{figure}

The over-temperature protection circuit is shown in Fig.~\ref{fig:sch_overtemp}.
Unlike other solutions that employ a differential or ratiometric temperature sensor to compensate for process variation and radiation effects~\cite{pertijs_2005,standke2019characterization}, 
the one adopted in ALDO2 uses the weak inversion "composite" MOS topology (MN1-2, red block)~\cite{Crepaldi12}, which provides a PTAT (proportional to absolute temperature) voltage that remains independent from any technology or working parameter (e.g., $\mathrm{V_{TH}}$, $\mathrm{I_{DS}}$, etc.) as long as MN1 and MN2 are both operated in the weak inversion region and neglecting second order dependencies on source-drain and source-bulk potential.
The PTAT voltage is defined by the following equation:

\begin{equation}
	\label{eq:vptat}
	\mathrm{V_{PTAT}} = \frac{kT}{q}\ln\left[\frac{\left(W/L\right)_{MN2}}{\left(W/L\right)_{MN1}}\right]
\end{equation}

The independence from technology parameters is anticipated to be advantageous for radiation hardness, as these parameters (e.g., $\mathrm{V_{TH}}$, carrier mobility) change as TID accumulates.

MN1 and MN2 are sized with a channel length $L$ of 1~$\upmu$m to minimize short-channel and mismatch effects, and an aspect ratio $W/L$ of 20 and  4980, respectively.
The devices are arranged in a matrix (25$\times$10), with MN1 at the center, further minimizing mismatches and thermal gradients.
The composite MOS device is biased by MP1 with 600~nA, resulting in a power consumption of approximately 1.1~$\mathrm{\upmu W}$ at 1.8~V supply.
The expected $\mathrm{V_{PTAT}}$ from Eq.~\ref{eq:vptat} is 143~mV at 27~$^\circ$C, with a thermal slope of 476~$\mathrm{\upmu V/^\circ C}$.
Simulations in the nominal corner give a PTAT voltage of 166~mV and a slope of 435~$\mathrm{\upmu V/^\circ C}$.
Monte Carlo simulations over process and mismatch give a spread of 1.2~mV RMS (0.7$\,\%$).

The PTAT voltage is then compared to an external threshold by a comparator (green block).
MP2 provides a hysteresis of 140~nA, which corresponds to 8.7~mV, or 18~$^\circ$C.
The over-temperature protection can be deactivated by pulling down the EN pin.

\begin{figure}
	\includegraphics[width=0.51\columnwidth]{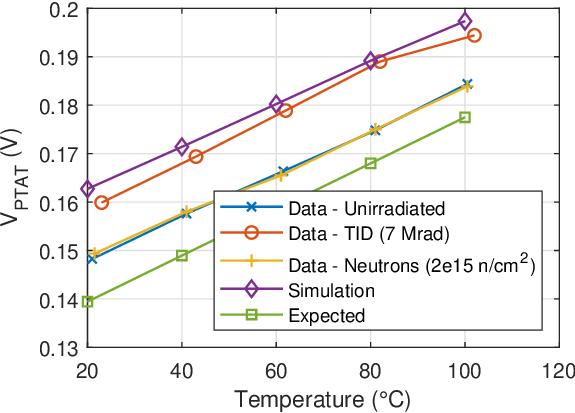}
	\hfill
	\includegraphics[width=0.48\columnwidth]{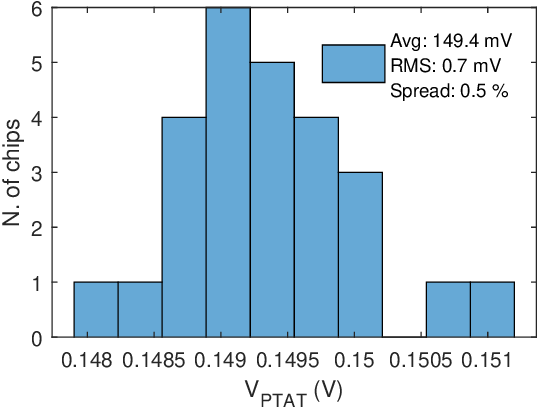}
	\caption{Left: Temperature sensor voltage ($\mathrm{V_{PTAT}}$) as a function of temperature for expected, simulated and measured chips. Right: Distribution of $\mathrm{V_{PTAT}}$ in 26 chips at ambient temperature.}
	\label{fig:overtemp_meas}
\end{figure}

Direct measurements of $\mathrm{V_{PTAT}}$ are not possible since the voltage is not available outside of the chip.
An indirect measurement was performed by operating the chip in a climatic chamber at different temperatures and decreasing the comparator threshold with a DAC, until the regulator would switch off.
This measurement, thus, includes offsets and other non-idealities of the comparator itself.
The results are shown in Fig.~\ref{fig:overtemp_meas}.
On the left, $\mathrm{V_{PTAT}}$ of expected, simulated, and measured (unirradiated and irradiated) chips is plotted versus temperature.
The measured slope is $(450\pm8)\,\mathrm{\upmu V/^\circ C}$.
Neutron irradiation (up to $\mathrm{2\cdot10^{15}\,n_{eq}/cm^2}$) does not affect the temperature sensor, while a 7~Mrad TID causes a shift of about 10~mV ($+22\,^\circ$C), but no change in the temperature slope.
On the right, an histogram of $\mathrm{V_{PTAT}}$ measured in 26 chips shows a good uniformity of 0.5$\,\%$ RMS, confirming the results of Monte Carlo simulations.

Thermal measurements with an infrared camera showed that protection kicks in at $105.4\,^\circ$C (comparator threshold at 189~mV) and then keeps the chip at $96.6\,^\circ$C, confirming that the hysteresis is $17.6\,^\circ$C (assuming a linear heating-cooling profile).

The excellent uniformity and only moderate sensitivity to TID prevent any need of calibration or adjustment of the threshold during the operation of the chip in the detectors.

\section{Layout and radiation hardening}
\label{sec:layout}

\begin{figure}
	\centering
	\includegraphics[width=\columnwidth]{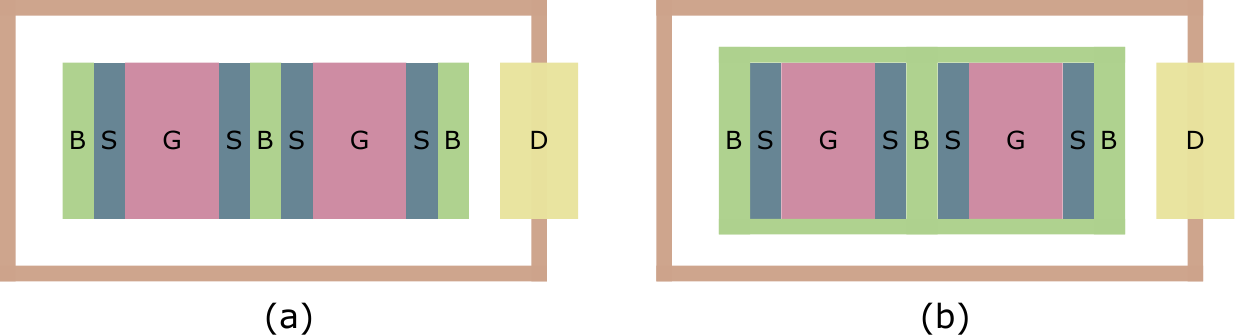}
	\caption{Top view of standard (a) and radiation hardened (b) layout of the HV vertical n-type DMOS. The bulk (P+) diffusion surrounds the device and prevents the thick oxide from facing the top and bottom edges of the source diffusions.}
	\label{fig:nmos_hv_layout}
\end{figure}

The I3T80 technology requires specific radiation hardening techniques in the layout of the chip.
All regular LV n-type transistors use the enclosed layout (annular gate)~\cite{anelli_1999}, where the channel does not interface with any thick oxide.
Moreover, the HV n-type vertical DMOSFETs adopt a modified layout, as illustrated in Fig.~\ref{fig:nmos_hv_layout}.
An additional P+ diffusion (bulk contact, in green) surrounds the device, preventing the thick oxide used by spacers and passivation from facing the top and bottom edges of the source.
This solution effectively reduces the leakage between drain and source resulting from the TID-induced accumulation of charges in the thick oxide~\cite{FaccioHVradhard}.

Additional hardening solutions implemented in the design of ALDO2 include the use of abundant well and bulk contacts, as well as guard rings, aimed at efficiently removing any charge deposited in the substrate and reducing the risk of single events effects (transients and latch-ups).
 
Being a power device, particular attention was also dedicated to controlling the maximum current density and minimizing the series resistance of high-current metal traces.
Traces were sized with large over-provisioning (over 3 times the I3T80 rules).
No power integrity software (e.g., Cadence Voltus) was used for this purpose.

Dummy structures and interleaving techniques are employed in all devices where matching is critical, except for DMOSFETs, as the penalty in area would be too large.

%
%
%

\section{Radiation hardness qualification}
\label{sec:radiation}

The chip's radiation hardness underwent comprehensive testing in several laboratories (including CERN and Karlsruhe for TID irradiation with X-rays, Ljubljana and Pavia for displacement damage irradiation with neutrons, and Legnaro for single-event effect irradiation with heavy ions).

The targeted radiation levels correspond to those encountered by the chip in the most demanding detector (CMS BTL), specifically 3.2~Mrad TID, $\mathrm{1.9\times10^{14}\,cm^{-2}}$ 1-MeV-equivalent neutron fluence, and $\mathrm{1.5\times10^{13}\,cm^{-2}}$ charged hadron fluence.

\begin{figure}[b]
	\centering
	\includegraphics[width=\columnwidth]{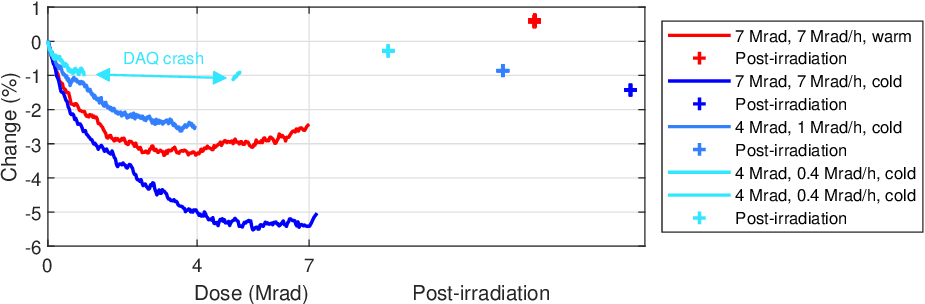}
	\caption{Online monitoring of the LV output during TID irradiation at different dose rates and temperatures. The red curve is taken at high dose rate (7~Mrad/h) and 30$\,^\circ$C. Blue curves are taken at 0.4, 1, and 7~Mrad/h at -30$\,^\circ$C. Post irradiation data (cross markers) are taken after 3 days at ambient temperature.}
	\label{fig:tid_rate_temp}
\end{figure}

The TID radiation hardness of certain CMOS technologies can be sensitive to irradiation temperature, especially at the low temperatures experienced in particle physics experiments, and dose rates~\cite{borghello2021tidtemp}.
Previous studies on onsemi I3T80 (made by CERN for the FEASTMP DCDC) confirm this observation for the technology employed by ALDO2.
Consequently, the X-ray irradiation for TID was conducted at both ambient and low temperatures, and varying the dose rate.

Fig.~\ref{fig:tid_rate_temp} presents the data of the online monitoring of the LV output during TID irradiation (using DTMOST-based bandgap) up to 4 and 7~Mrad.
A comparison between the red data (7 Mrad/h, 30$\,^\circ$C) and the dark blue data (7 Mrad/h, -30$\,^\circ$C) clearly reveals that the output voltage change is significantly greater (almost double, 5$\,\%$ vs 3$\,\%$) for the chip irradiated at low temperature.
A similar difference persists even after annealing at ambient temperature for 3 days (cross markers).
The impact of dose rates at fixed temperature (blue data, from 0.4 to 7~Mrad/h) is even more pronounced, with lower dose rates resulting in less output voltage change ($<$1$\,\%$) both during irradiation and after annealing.

TID irradiation does not significantly affect the temperature drift of the bandgaps.

The effects of TID on the HV regulator are similar to the LV one, but the change is smaller ($<$0.2$\,\%$).
The OCM circuit within the HV section is the most affected part, with a measurement change up to 10$\,\%$ at 7~Mrad (twice the maximum radiation level expected in the working environment).
However, as discussed in Sec.~\ref{sec:lv_meas}, the OCM circuit can be easily re-calibrated by measuring the output current through the feedback network.

\begin{figure}
	\centering
	\includegraphics[width=0.9\columnwidth]{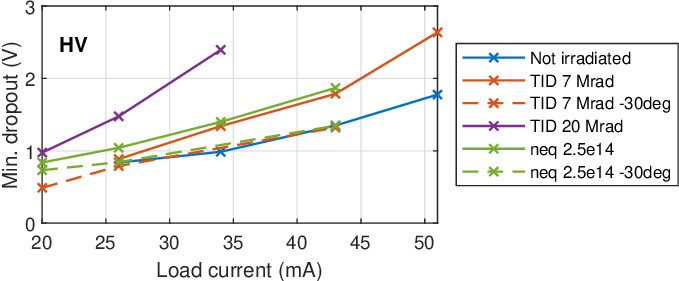}
	
	\vspace{0.2cm}
	\includegraphics[width=0.9\columnwidth]{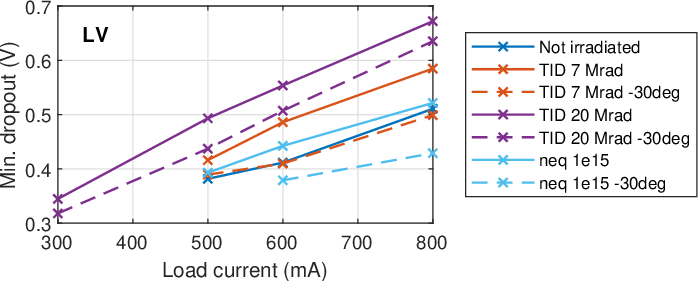}
	\caption{Minimum dropout as a function of load current of the HV regulator (top) and LV regulator (bottom) for chips irradiated with X-rays and neutrons. Dashed lines are measurements performed at the operating temperature (-30$\,^\circ$C).}
	\label{fig:dropout_rad}
\end{figure}

The most relevant effects of both TID and neutron irradiations on ALDO2 concern the line regulation and the minimum dropout.
Both types of irradiation diminish the output current capability of both the HV and LV regulators at a given input voltage due to an increase of the on-resistance of the pass transistor, necessitating the use of higher dropout to maintain adequate line regulation and supply rejection.
Fig.~\ref{fig:dropout_rad} shows the minimum dropout as a function of load current at different radiation levels and working temperatures.
In this test, all irradiations were conducted at ambient temperature.
Operating the irradiated chips at low temperature (dashed data) allows for the nearly complete restoration of the performance of the unirradiated chip (solid blue data) for both the HV and LV regulators, with the exception of the chip irradiated at 20~Mrad.
The impact of displacement damage from a neutron fluence of $\mathrm{10^{15}\ n_{eq}/cm^2}$ (5 times above the expected levels) results in an increase in the on-resistance of the pass element of the HV regulators, reaching up to 500~$\Omega$. This increase subsequently limits the output current to a maximum of 10~mA.

Another part of the ALDO2 that is potentially sensitive to radiation damage is the VBIASH and VBIASL internal generation circuit described in Sec.~\ref{sec:hv_circuit}, which makes use of diode-connected MOSFETs sensitive to threshold voltage drifts.
There is indeed a significant drift of the voltages generated by this block, up to 700~mV, but this remains low enough to ensure the operation of ALDO2 at the radiation levels required and with an input voltage range of approximately 1:1.6 (e.g., 32-50 V).

Among the three bandgaps included in the ALDO2, the DTMOST-based one was deemed the most suitable, as it offers the best compromise between thermal stability and radiation hardness, with changes below 1.5$\,\%$ at the target radiation levels.
Conversely, the PNP-based one exhibits a drift of up to 4$\,\%$ with neutron irradiation and 2$\,\%$ with X-rays.

\begin{figure}
	\includegraphics[width=0.485\columnwidth]{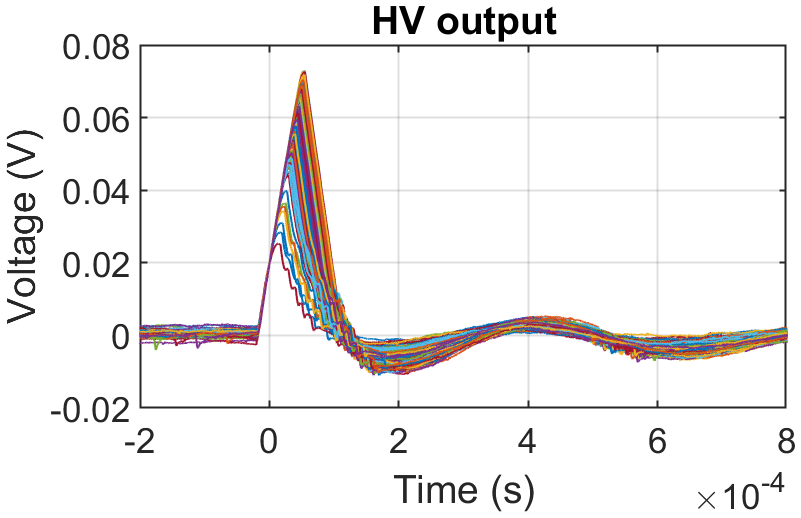}
	\hfill
	\includegraphics[width=0.485\columnwidth]{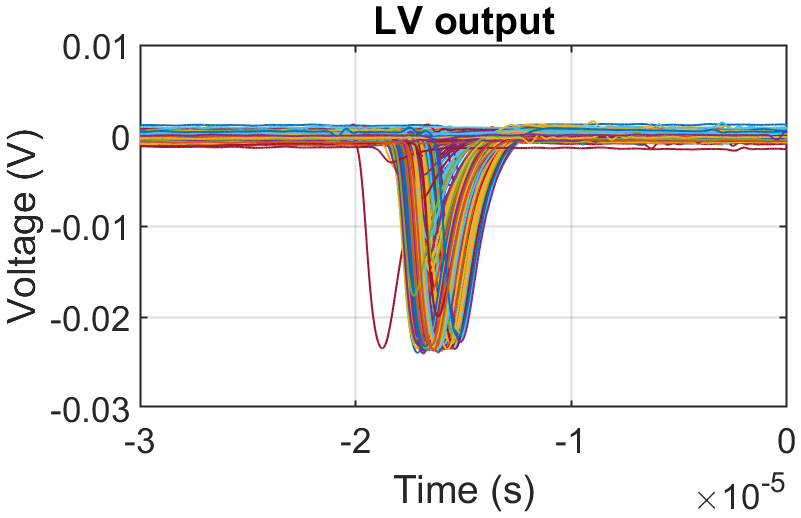}
	\caption{Data acquisitions with an oscilloscope of SETs. The trigger was set on the HV output (level 20~mV). SETs are synchronous between HV and LV regulator approximately 50$\,\%$ of the triggers. SETs are always synchronous between the two HV regulators.}
	\label{fig:transients}
\end{figure}

Finally, single events effects (SEE) were studied with heavy ion irradiation.
No single-event latchup (SEL) was found after a cumulative fluence of $\mathrm{1.7\cdot10^{10}\ ions/cm^2}$ with LETs between 8.6 and 54.7~$\mathrm{MeV\,cm^2/mg}$. This allowed for the computation of a SEL cross-section upper limit of $\mathrm{2.3\cdot10^{-9}\ cm^2}$ at a LET of 8.6~$\mathrm{MeV\,cm^2/mg}$ and $\mathrm{1.2\cdot10^{-10}\ cm^2}$ at 28.4~$\mathrm{MeV\,cm^2/mg}$ (following ESA-ESCC 25100 guidelines with 68$\,\%$ C.L.).

Single-event transients (SET) were observed, as shown in Fig.~\ref{fig:transients}.
The sensitive circuit was identified in the over-temperature protection, which triggers synchronous transients in the three protected regulators (the HV ones and the main LV one).
The transients exhibit moderate amplitude and duration (70~mV and 150~$\upmu$s on the HV output, 22~mV and 5~$\upmu$s on the LV output).
Introducing an external filtering capacitor on the comparator threshold significantly reduced the rate of SETs by almost two order of magnitude.
The cross-section with this mitigation in place is $\mathrm{2.1\cdot10^{-7}\ cm^2}$ at a LET of 28.4~$\mathrm{MeV\,cm^2/mg}$.
Due to limited beam time availability, it was not possible to measure the full Weibull curve of the SET cross section as a function of LET.
Nevertheless, even assuming a very conservative threshold LET of 1~$\mathrm{MeV\,cm^2/mg}$, the SET cross section with protons would be $\mathrm{4.6\cdot10^{-12}\ cm^2}$ (computed from the heavy ion cross-section as discussed in \cite{edmonds2000seu}),
translating to 10.3 SETs per year per chip in the region of the BTL detector closest to the endcaps.
Given the amplitude and duration of the observed SETs, this amount is well within acceptable limits.

\section{Production yield, reliability and chip testing}

The ALDO2 chip was designed complying with most of the design-for-manufacturing (DFM) and all of the DRC rules suggested by onsemi in order to maximize the yield.

A random sampling of the devices from 8 of the 26 wafers allowed for the estimation of a high yield of 99.5$\,\%$ (98.7-99.8$\,\%$ with C.L. 90$\,\%$~\cite{birolini1999reliability}).
It should be noted that the yield measurement was conducted using a chip socket, which could damage the chip if critical pins are not properly contacted, potentially worsening the measured yield.

The aging of the chip was tested in an accelerated environment, involving 700 hours at 90$\,^\circ$C and 240 hours at -55$\,^\circ$C, with no failures observed in the tested sample.

The exceptional yield and reliability of ALDO2, along with the very high uniformity demonstrated in Sec.~\ref{sec:hv_meas} and \ref{sec:lv_meas}, have made it possible to skip individual chip testing prior to PCB mounting, as there is no need for individual trimming, neither across wafers.

\section{Summary and future prospects}
\label{sec:summary}

This article presented the ALDO2 ASIC, a multi-function, radiation-hardened power management solution designed for SiPM-based particle physics detectors.

The chip has been manufactured in the required quantities for the BTL and HGC detectors of the CMS experiment at CERN (45k chips) and, at the time of writing this article, is about to being assembled on the final PCBs of the front-end boards.

Due to ALDO2 unique features and characteristics (e.g., operation at 60~V, radiation hardness above few Mrad of TID and $10^{14}\ \mathrm{n_{eq}/cm^2}$ of neutron fluence, etc.), direct comparison with other state-of-the-art solutions is not feasible.

Considering the anticipated increase in the utilization of SiPMs in particle physics experiments, there exists the potential for further enhancements of the chip, incorporating additional functionalities.
The design of ALDO3 should initiate with the selection of a different technology, since I3T80 from onsemi is becoming outdated and the fabrication facility used for the production of the chip and qualified for radiation hardness has ceased operations.
Valuable improvements to the chip design could include the integration of digital circuits (e.g., DAC, ADC, I-V curve measurement logic, etc.), as well as the adoption of a capacitor-less compensation scheme for the HV regulators, which would circumvent the large PCB area currently occupied by HV tantalum capacitors.

\appendices


\section*{Acknowledgments}

The authors would like to thank IMEC and Europractice for their invaluable support, F. Faccio at CERN for the suggestions regarding the I3T80 technology, and all the personnel at the irradiation facilities (CERN, Karlsruhe, Legnaro, Ljubljana, and Pavia).
The authors also thank the colleagues from CMS BTL and HGC collaborations, which supported and funded the development of the chip.
A generative AI tool (Opera's Aria) was employed as a proofreading tool to enhance the English language quality of the paper.

\bibliography{main}

\end{document}